# Many-Body Configurational Spectral Splitting between Trion and Charged Exciton in a Monolayer Semiconductor

Jiacheng Tang,[1,2] Cun-Zheng Ning[2,1]*

[1]Department of Electronic Engineering, Tsinghua University, Beijing 100084, China
[2]College of Integrated Circuits and Optoelectronic Chips, Shenzhen Technology University, Shenzhen 518118, China

* Corresponding author. Email: ningcunzheng@sztu.edu.cn

## Abstract

**Many-body electron-hole complexes in a semiconductor are important both from a fundamental physics point of view and for practical device applications. A three-body system of electrons (e) and holes (h) (2e1h, or 1e2h) in a two-band semiconductor is commonly believed to be associated with two spectral peaks for the exciton and trion (or charged exciton), respectively. But both the validity of this understanding and the physical meaning of a trion or charged exciton have not been thoroughly examined. From the physics point of view, there are two different configurations, <e><eh> or <eeh>, which could be considered charged exciton and trion, respectively. Here <···> represents an irreducible cluster with respect to Coulomb interactions. In this paper, we consider these issues related to the 2e1h three-body problem theoretically and experimentally using monolayer MoTe$_2$ as an example. Our theoretical tools involve the three-body Bethe-Salpeter Equation (BSE) and the cluster expansion technique, especially their correspondence. Experimentally, we measure the photoluminescence spectrum on a gate-controlled monolayer MoTe$_2$. We found two spectral peaks that are 21 and 4 meV, respectively, below the exciton peak, in contrast to the single "trion" peak from the conventional understanding. We show that, while the three-body BSE in a two-band model can reproduce all spectral features, the cluster-expansion technique shows that the two peaks correspond to the charged exciton <e><eh> and trion <eeh>, respectively. In other words, there is a spectral splitting due to the two different many-body configurations. Furthermore, we find that the trion only exists in the intervalley case, while the charged exciton exists both for the intervalley and intravalley cases. Importantly, we point out that the new spectral feature is a pure many-body splitting and should not be confused with the fine structure of trion, usually associated with the spin-split bands and ee/eh exchange interaction. Additionally, our theory also could explain similar spectral features in previous experiments on MoSe$_2$, demonstrating the universality of the many-body configurational splitting. Our results provide a new understanding and a more complete picture of three-body (and other many-body) systems.**

**Main**

Understanding the formation mechanism, different possibilities of configurations, and basic features of various quasi-particles is not only important for basic many-body physics in condensed matter, but also for many technologically important photonic applications. 2D materials provide an important platform for both purposes. Even though excitons and trions have been widely studied in 2D materials and other low-dimensional materials, and in conventional semiconductors, some of the very basic questions remain unclear, unanswered, or even unasked in certain cases, especially for TMDCs. In the case of a three-body system with two electrons (e) and one hole (h) (2e1h, or its charge conjugate counterpart, 1e2h) in a two-band semiconductor, the conventional absorption or emission spectrum is thought of containing two spectral peaks for exciton (X) and trion (T) (or often interchangeably called charged exciton), respectively. But both the validity of this picture and the question of what exactly is a trion or charged exciton have not yet been thoroughly examined. From the cluster expansion point of view, a 2e1h system can allow two different configurations, <e><eh> or <eeh>, where <...> represents an irreducible cluster with respect to Coulomb interactions. Intuitively, the <e><eh> configuration is associated with charged exciton, while <eeh> represents an irreducible cluster of three bodies. More importantly, are these two configurations spectrally distinct or identical? These and the related questions are the focus of this paper.

The issue of different many-body configurations or multiplicity of bound states is a more general one and occurs in many different contexts. In the field of ultracold atomic gas or condensed matter, for example, the issue manifests as Feshbach resonance or the Efimov Effect[1,2]. In the Mott transition physics, the diversity of configurations in three-body (3B) (trion[3-10]) and four-body (4B) case (such as bi-exciton[11-16] or quadron[17]/quadruplon[18]) leads to complex mixtures of different species in an intermediate density region, deviating from the simple picture of exciton-plasma transition. In the 1D Hubbard model theory[19], triplon, quadruplon, and multiplon associated with the irreducible clusters of the corresponding orders were found as bound states in the three-, four-, and N-hole energy loss spectra, respectively. It is worth noting that a recent theoretical study has pointed out the similar difference between trion and charged exciton for a confined quantum dot system based on the Hartree-Fock method[20]. Our concern in this paper is a much more general understanding of a generic three-body 2e1h or 1e2h system. In general, searching for possible new bound states corresponding to different many-body configurations is always one of the essential tasks of many-body physics.

In this letter, we attempt to answer the above questions by conducting a combined theoretical and experimental investigation into the possible bound states of a 2e1h 3B system in monolayer $MoTe_2$. One important advantage of $MoTe_2$ (also to a large extent, $MoSe_2$) is the large splitting of the spin states in the conduction band, such that the higher spin band can be ignored within the small energy range of interest in this paper. Therefore, no other spin configurations can cause trion splitting as studied in other single-layer TMDC materials[21-23]. We therefore treat a two-band problem only. Theoretically, we solve the three-body Bethe-Salpeter Equation[21-28](3B-BSE) to obtain the optical spectrum. We then recast the Feynman diagrammatic representation of the 3B-BSE into the cluster expansion picture. We show that the 2e1h system has two different configurations <e><eh> and <eeh> with two distinct spectral peaks. While the former corresponds to an exciton weakly coupled to an electron (thus resembling more a charged exciton), the latter corresponds to a three-body irreducible cluster, or a triplon using the language of Ref. [19]. Furthermore, the correspondence between the two approaches allows us to associate different many-body configurations to different spectral peaks. Experimentally, we performed photoluminescence (PL) spectroscopy under pumping of a continuous-wave laser in monolayer $MoTe_2$. Interestingly, our experimental measurements completely verified the above theoretical prediction with two peaks that are 21 meV and $3 - 4$ meV below the exciton peak, in quantitative agreement with the theoretical results. To further verify the spectral features and the new physical picture of understanding of the three-body problem, we performed the same calculation for another 2D material, $MoSe_2$, which has similar band structure to that of $MoTe_2$. Our theoretical calculation produced similar spectral feature, which is in quantitative agreement with the experimental results of an earlier paper by other group[8], demonstrating the universal character of the new three-body physics.

**Three-body Bethe-Salpeter Equation and the cluster expansion technique**

In quantum field theory for many-body systems, the well-known Bethe-Salpeter Equation[29-31] (BSE) is used to study quasi-particles such as diquark, meson, and exciton, etc, within the 2B interaction, called 2B-BSE in this letter. In the 3B case, 3B-BSEs[21-24,26,27] or the 3B version of the Semiconductor Bloch Equations[25,28] (SBEs) were developed to calculate the energies of those 3B systems, such as baryon, proton, neutron, and trion, etc. With the recent advent of low-dimensional nanostructures such as 2D materials, 3B-BSE has attracted renewed interest in the calculations for the 3-particle bound energies[24,25], fine structures[21-23], cavity-coupled polaritons[32], and the optical spectra[26-28].

Similar to 2B-BSE, 3B-BSE takes also the form of an integral equation, of which the Feynman diagrammatic representation is shown in Fig. 1(a). Usually one needs to solve such an integral equation to obtain the 3-particle Green's function (green part). Various integral (interaction) kernels (blue dashed box) are specified in Fig. 1(b), including e-e and e-h interactions (top row) and their Fermi permutation terms (bottom row) for a 2e1h system. For the sake of brevity, we do not present the e-h exchange terms explicitly in Fig. 1(b).

It is known that 2B-BSE could be always approximated by a sum of ladders, or called ladder approximation[33,34], to correct the vertex or to expand the correlation function via the 2-particle interaction kernels. In our 3B case, the 3B-BSE has essentially also the form of self-consistent iteration. One could replace the green part on the right in Fig. 1(a) with the entire right side of the equal sign of the expression, and repeat it indefinitely. In this way, the 3-particle correlation function could be expanded by such infinite iterations into the sum of a series of 3-propagator ladders, as shown in Fig. 1(c).

The kernels in Fig. 1(b) connect the three propagator lines and stack in various combinations. Such combinations are classified into different partial summations, which can be associated with the irreducible clusters of different orders that we will discuss in the following.

Apart from BSE, cluster expansion is another frequently-used approach to describe an interacting system. As we know, the cluster expansion approach was successfully formulated to study the hierarchy problem of those of many-body physics or semiconductor quantum optics[35,36], such as e-h dropleton[37] and the quantum optical spectroscopy[35,36], etc. Recently the cluster expansion approach was combined with the Semiconductor Bloch Equation to model the excitonic Mott transition[28] in an intravalley case for $MoS_2$, up to the second order irreducible cluster. But for a more general 2e1h/1e2h-type 3B problem with the intervalley interactions, the cluster expansion has not been fully investigated up to the third order irreducible cluster.

In Fig. 1(d), we write down the complete sequence of the cluster expansion for a 2e1h system. The specialization of such cluster expansion for ML-$MoTe_2$ yield the similar sequence of clusters, now illustrated together with the spin-valley-resolved band structures in Fig. 1(e). In both Fig. 1(d) & 1(e), $\triangle$ represents an irreducible cluster of the $n^{th}$-order with n Fermions. In this way, $\triangle$ is a quasi-free electron or hole. $\triangle$ represents a direct or indirect *e-h* pair, or 2-body (2B) state in general. $\triangle$ represents a 2*e*1*h* 3-body (3B) irreducible cluster. In the language of Ref. [19], $\triangle$, $\triangle$, and $\triangle$, are singlon, doublon, and triplon, respectively.

As can be seen in Fig. 1(c), the different partial summations do correspond to the above irreducible clusters of various orders. For examples, the partial summations where one of the two electron lines is connected with the hole line while the other one is free (the second and third row of the expression in Fig. 1(c)), are clearly equivalent to the two configurations of ⚠ ⚠ (considering the indistinguishability of the two electrons). The partial summation that all the three lines are connected (the fourth row of the expression in Fig. 1(c)) is equivalent to the 3B irreducible cluster, ⚠. In this way, the correspondence can be established between the BSE and the cluster expansion picture for a 3B system. Similarly, such correspondence for a more complicated 2e2h 4B system can also be established[18].

The infinite hierarchy of the cluster expansion is often truncated at a finite order for practical calculations. The effects of the higher orders introduce couplings among the lower orders of irreducible clusters, or effective interactions among multiplons of various orders. Such effective interactions are indicated by the wavy lines in Fig. 2(c). In this letter, we truncated the cluster expansion at the 3$^{rd}$ order. As a result, the original non-interacting cluster ⚠ ⚠ shown in Fig. 1(c) – 1(e) becomes weakly interacting. We indicate such weak coupling by adding a wavy line between the irreducible clusters, *i.e.* ⚠ ~ ⚠ (see Fig. 2(d)). Obviously, cluster ⚠ or triplon represents completely a distinct physical entity from ⚠ ~ ⚠. The most fundamental difference is the absence of an exciton in ⚠ and the lack of a clear association of any one of the two electrons to the hole. From the above description, it seems more appropriate to name cluster ⚠ "trion" and to name ⚠ ~ ⚠ "charged exciton". In this way, "trion" and "charged exciton" are completely two different entities of a 3B system. Since the term "trion" has been used in literature interchangeably with "charged exciton" as also pointed out by Quang and Huong[20], one could adopt the term of "triplon" from Ref. [19] or "trion" to refer to the irreducible cluster of ⚠, and leave the conventional term "charged exciton" or "doublon-singlon" for the weakly coupled ⚠ ~ ⚠.

It is important to point out that, the spectral splitting between ⚠ and ⚠ ~ ⚠ is fundamentally distinct from the trion fine structure in literature[21-23]. The trion fine structure found in WS$_2$, WSe$_2$, and MoS$_2$ always involves two conduction bands, while the spectral splitting discussed here is induced by the two many-body configurations and requires only the lower-energy conduction band. This is also why the trion fine structures have not been observed for MoTe$_2$ or MoSe$_2$[8] due to the order of the two conduction bands (bright exciton being the lower energy band) and the large separation between the two conduction bands. The configuration related splitting discussed here would still exist even if we exclude the e-h and e-e exchange terms, while the trion fine structure would degenerate to a single trion state.

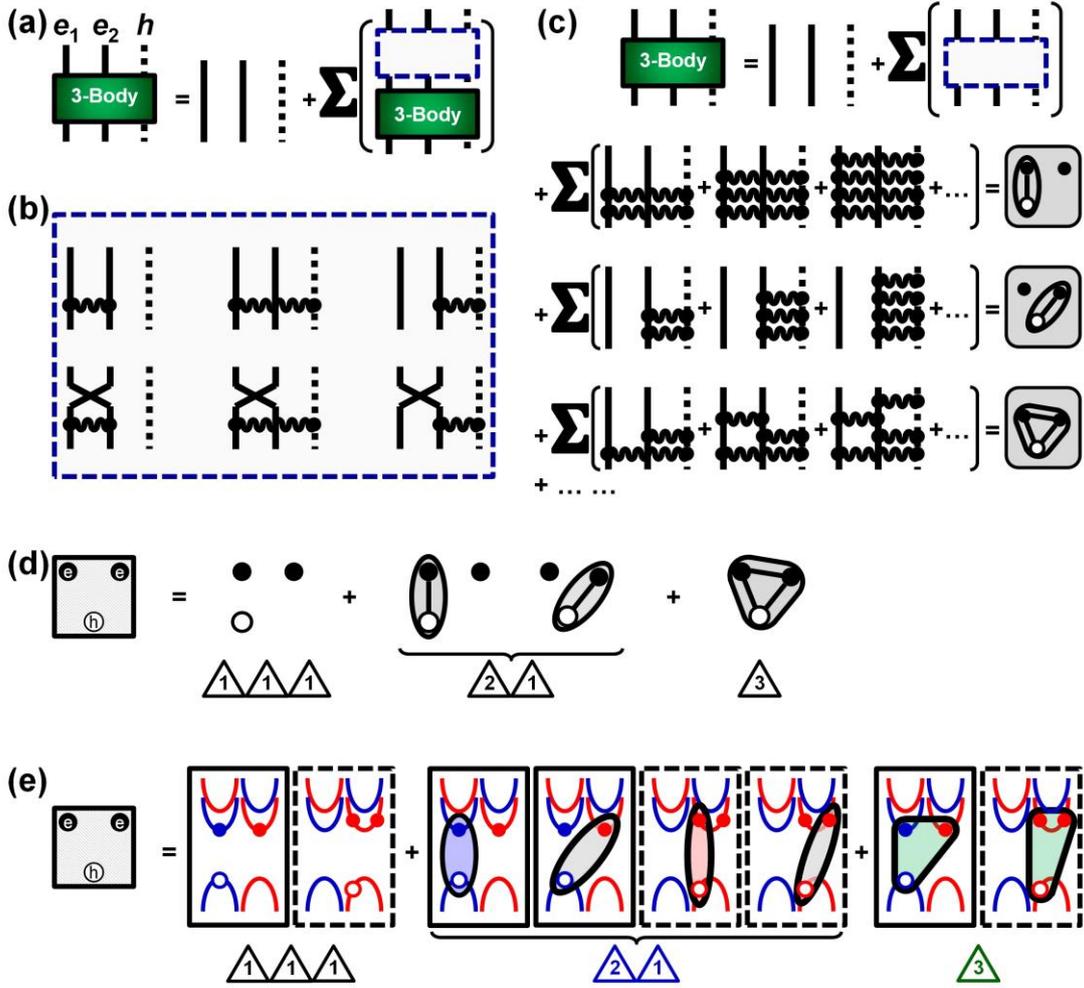

**Fig. 1 (a),** Feynman diagrammatic representation of the 3B BSE. **(b),** Feynman diagrams of the interaction kernels. These kernels should be inserted in the blue dashed box in **(a)** and summed over all together. For the sake of brevity, the positive or negative signs of these kernels are not specified. The e-h exchange terms are not explicitly illustrated. **(c),** Feynman diagrammatic representation of the 3B BSE in **(a)** expanded into the summation of infinite series. The Fermi permutation terms (the bottom row of **(b)**) are omitted in **(c)** for brevity. **(d), (e),** Schematic of the cluster expansion of the 2e1h 3B system **(d),** and the analogous ones shown together with the spin-valley-resolved band structure of ML-MoTe₂ **(e)**. The 2e1h 3B system is expanded into irreducible clusters of various orders. In **(e)**, the spin-up and spin-down single-particle states are colored in blue and red, respectively. The intervalley 3B states are shown with solid-line boxes, while the intravalley ones are shown with dashed-line boxes. It is interesting to point out that the two △

△ configurations with dashed boxes are quantum-mechanically indistinguishable. For brevity, we exclude those 2B clusters with the same charges, such as e-e, in **(c)** − **(e)**, but these clusters are included in our theoretical calculations.

Even though the above physical picture and the theoretical treatment are valid for two-band semiconductors in general, we would like to specialize the above theory to the case of MoTe$_2$ to calculate its optical transition spectrum, with the results shown in Fig. 2(a) & 2(b). Fig. 2(c) shows the optical transition in correspondence with the scale of the total 3B energy spectrum. Suppose the VBM to be the zero-energy point, the 3B entities have the total 3B energies of 2.914, 2.931, & 2.934 eV, respectively. Referring to the total energy of an exciton-electron continuum (XeC): $E_{1s-X} + E_e \approx 1.770 + 1.164 = 2.934$ eV, we could confirm the existence of two major bound states in total, of which the spectral peaks of optical transitions are located at 1.144 (P$_1$) and 1.161 (P$_2$) eV, i.e. ~20 (P$_1$) and ~3 − 4 (P$_2$) meV below X, as shown in Fig. 2(b), respectively.

To examine the roles played by different clusters in the spectrum, we calculate optical transitions by truncating the cluster expansion to different orders. For truncation up to △ △, the spectrum is shown in Fig. 2(a), where we see a single spectral line at the location of the exciton. The truncation up to △ △ indicates the absence of any interaction between the exciton and the electron, thus producing the only single X peak in Fig. 2(a). By solving the full 3B-BSE with △ included, we could see two important changes in Fig. 2(b) compared with Fig. 2(a): i) P$_1$ is newly produced; ii) the original X splits into two peaks: P$_2$ & P$_3$. The emergence of P$_1$ indicates the effect of cluster △ or the triplon. While the splitting of X implies the existence of another bound state P$_2$ that has a binding energy of 4 meV relative to X, due to the coupling between △ and △ through the inclusion of a higher order, i.e. △. Or equivalently, the irreducible cluster △ converts △ △ to a weak-coupling △ ~ △. In this sense, our results so far show that charged exciton, △ ~ △, and trion (or triplon), △, are physically and spectrally distinct.

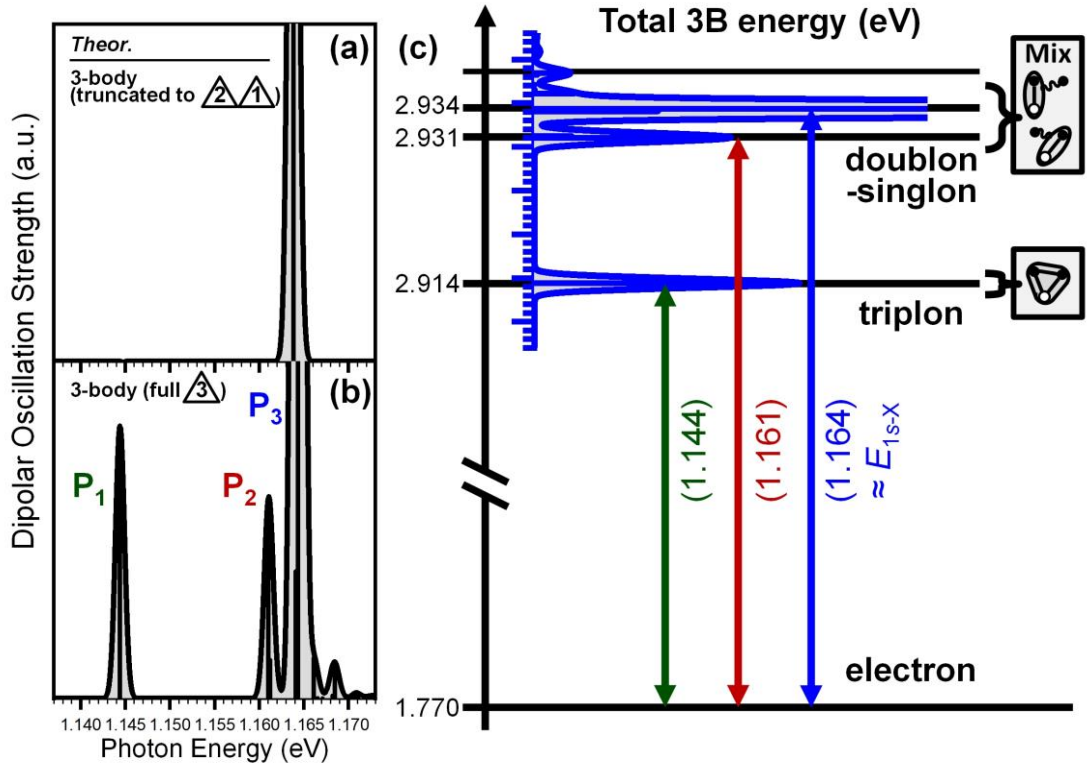

**Fig. 2 (a), (b),** Theoretical spectra calculated for ML-MoTe$_2$ based on the 3B-BSE. **(a) & (b)** show the spectra of transition between the electron and the 3B states obtained from the 3B-BSE truncated up to △△△ **(a)** and △ **(b)**, respectively. **(c),** Energies of different states calculated from the 3B-BSE, with the corresponding clusters identified through the cluster expansion approach. The transitions corresponding to the spectrum in **(b)** are also marked.

## Experimental results

To experimentally verify the theoretical results and the associated physical picture of understanding presented above, we perform photoluminescence experiments on a gate-controlled ML-MoTe$_2$. We fabricated samples with a ML-MoTe$_2$ sandwiched between two hexagonal boron nitride (h-BN) layers and a back gate was used to control the background charge (see Ref. [7,9,10] for the schematic of the device structure). Fig. 3(a) – 3(c) show typically the photoluminescence (PL) spectra with the pumping of a 633-nm continuous-wave (CW) laser at 4K at different gate voltages ($V_g$) with the corresponding doping levels of charge carriers. Fig. 3(d) maps the PL spectra by sweeping $V_g$ with small steps. It is seen from Fig. 3(a) – 3(d) that there are mainly two regions of spectral features marked by X and T (T$^-$ or T$^+$). These are spectral peaks observed in typical gate-dependent measures of 2D materials that are identified as trion (peak T) and exciton (X), respectively. However, a closer examination of features around X shows that this feature contains more than one peaks reflected in the asymmetric shape of it, as shown in Fig. 3(a) – 3(c) for three specific gate voltages more clearly where we also performed multi-Gaussian-peak fitting, as shown by the green, red, and blue peaks. The X feature clearly contains two peaks separated by 3 – 4 meV from our fitting. The second-order differential (SOD) is commonly used to analyze spectral features in similar situations[38]. The SOD mapping, which corresponds to the second order derivative of Fig. 3(d), is shown in Fig. 3(e), showing more clearly the spectral splitting around feature X. Furthermore, we extracted the maximum values of the mapping in Fig. 3(e) and show in Fig. 3(f). We now see clearly that there are altogether 3 peaks, marked by $P_1$, $P_2$, & $P_3$, contrary to the common picture of two peaks corresponding to exciton and trion. We also show comparison of the peaks extracted from the direct SOD in Fig. 3(a) – 3(c) (lower panels) with the result of multi-peak fitting and found excellent agreement, establishing the robustness of peaks $P_2$ & $P_3$. Fig. 3(g) & 3(h) show the integrated areas and the FWHM of $P_1$, $P_2$, & $P_3$, extracted from the fitting, respectively, plotted as functions of $V_g$.

In Fig. 3(e) & 3(g), we see $P_1$ exist in the doping regime but becomes weaker around the charge-neutral regime (CNR), while both $P_2$ & $P_3$ only exist around the CNR or in the low-doping regime (LDR) but disappear in the high-doping regime (HDR). As can be seen in Fig. 3(h) as well, the linewidth of $P_1$ decreases with increasing deviation from charge neutrality, while $P_2$ & $P_3$ obviously broaden. Such broadening leads to the overlapping of $P_2$ & $P_3$ in the HDR, which makes it difficult to distinguish them by either fitting or analyzing the SOD. It could be seen also in Fig. 3(f) that from the CNR to HDR $P_1$ shows an obvious red-shift, while $P_2$ & $P_3$ do not seem to shift around the CNR and in the HDR with increase in the doping density the overlap of $P_2$ & $P_3$ shows a blue-shift. This implies $P_1$ could be a strongly charged entity, while $P_2$ & $P_3$ could be weakly charged entities. As we know, the 3B model is quite applicable for LDR but not for HDR. For HDR, excitons are dynamically dressed by the Fermi sea where the concept of trions becomes invalid and the system is more appropriately described by exciton polaron[39,40].

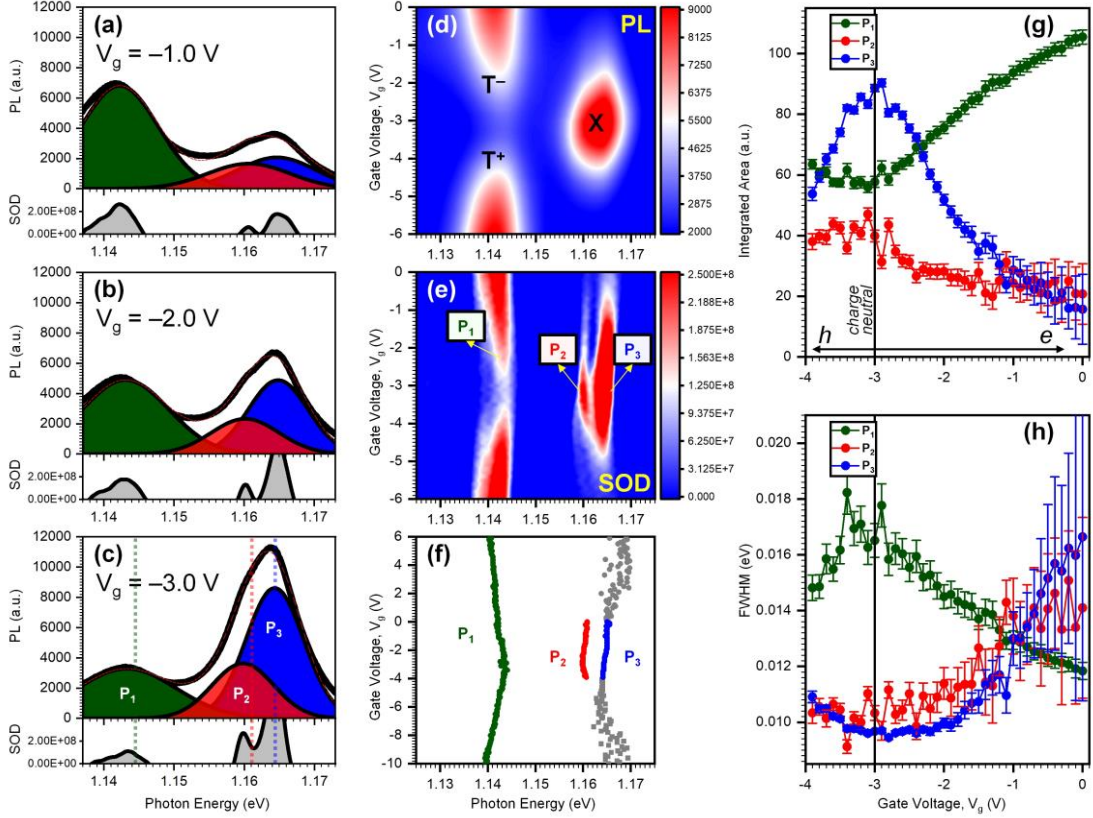

**Fig. 3 (a), (b), (c),** Upper panel: PL spectrum (black solid line) and the fitting curve (red dashed line) with three Gaussian peaks, $P_1$, $P_2$, & $P_3$; Lower panel: SOD (second order derivative) result of the PL spectrum of the corresponding upper panel. **(d),** PL mapping in the plane of photon energy and gate voltage, $V_g$. **(e),** SOD performed for **(d)** over the photon energy. **(f),** plot of the energy position of the extreme from **(e)** as a function of $V_g$. The results in **(a), (b), & (c),** are the representative profiles of **(d) & (e)** at $V_g$ = −1.0 V **(a)**, −2.0 V **(b)**, & −3.0 V **(c)**, respectively. **(g), (h),** Plots of the integrated area **(g)** and the full width at half maxima (FWHM) **(h)** of the three fitting peaks as functions of $V_g$. In **(g) & (h)**, the charge-neutral point is marked by the vertical line at $V_g$ = −3.0 V. The regimes of $V_g$ < −3.0 V and $V_g$ > −3.0 V correspond to the conditions of hole-doping (*h*) and electron-doping (*e*), respectively.

To show quantitative comparison between our theoretical results with experimental measurement, the three spectral lines calculated from theory (Fig. 2(b)) are over-laid on top of Fig. 3(c). Clearly, we see good agreement is found with differences around 1-2 meV.

To show the general validity of the new physical picture and the theoretical calculation, we also calculated similar 3B spectra for ML-MoSe$_2$, and compare the theoretical results (Fig. 4(b)) with experimental spectra (Fig. 4(a)) measured by Arora et al[8]. As can be seen from Fig. 4, the features of $P_1 - P_3$ from our 3B theory are in good agreement with those fitted from the experiments for ML-MoSe$_2$ where the error for each peak's resonance is less than 2 meV.

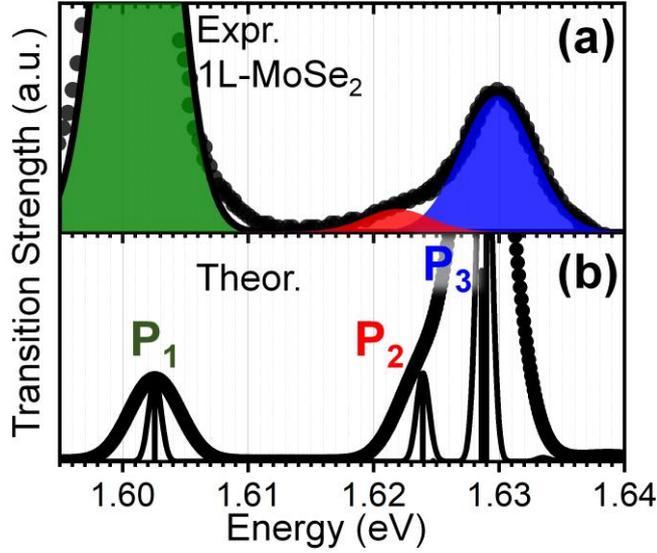

**Fig. 4 (a),** Experimental PL spectra at 5K for ML-MoSe$_2$ reproduced from Ref. [8]. We performed similar multi-peak fitting to that in Fig. 3(a) – 3(c), with the results represented by the filled green, red, and blue profiles. **(b),** Theoretical spectra calculated for ML-MoSe$_2$ based on our 3B-BSE.

To shed more light on the spectral peaks $P_1$ and $P_2$, we calculated the 1B-3B transition spectra separately for the intervalley and intravalley contributions, with the results shown in Fig. 5(a) & 5(c) and Fig. 5(b) & 5(d), respectively, where we also tested the numerical convergence by changing grid sizes (Fig. 5(a) & 5(b)) and the cut-off wavevector (Fig. 5(c) & 5(d)). The intervalley and intravalley 3B configurations are shown in the solid- and dashed-line frames, respectively, in Fig. 1(e). One can see they correspond to the two situations where the e-h pair is located at the same valley or not with the other electron. We note that the two electrons of the intravalley 3B state are in different momentum states and thus satisfy the Pauli principle. Comparing Fig. 5(a) & 5(c) and Fig. 5(b) & 5(d), we could see $P_1$ (corresponding to the irreducible cluster of order 3) only exist for the intervalley case but not for the intravalley one. This implies that the intravalley triplon does not exist, of which the two electrons have the same spin and orbital angular momentums (SAM or OAM). While the intervalley triplon could exist due to the different SAM or OAM of the two electrons. It seems quite analogous to the Cooper pairing of two electrons with opposite spins or the bounding and anti-bounding states of a Hydrogen molecule. In Fig. 5(a) – 5(d), $P_2$ & $P_3$ exist for both of intervalley and intravalley cases. This implies that even if the two electrons have the same SAM or OAM, the 3B irreducible correlation could always induce the renormalization (wavy line) among △ △ although it does not contribute to the 3B irreducible entity. Incidentally, our 3B-BSE results for the intravalley case agree with the previous calculations[28] based on SBE that did not include the intervalley case.

To check the numerical convergence, we show the spectra calculated with different k-mesh sizes (Fig. 5(a) & 5(b)) and the truncation wavevectors (Fig. 5(c) & 5(d)). In addition, we plotted the spectral positions of $P_1 - P_3$ from Fig. 5(a) – 5(d) with respect to the mesh sizes (Fig. 5(e) & 5(f)) or k-cutoff (Fig. 5(g) & 5(h)). It is clear that the calculated results of $P_1 - P_3$ is numerically stable and systematically converging. We found that all the spectral peaks converge reasonably well at the k-grid size of ~60×60 ×1 and at the k-cutoff of ~0.170 Å$^{-1}$.

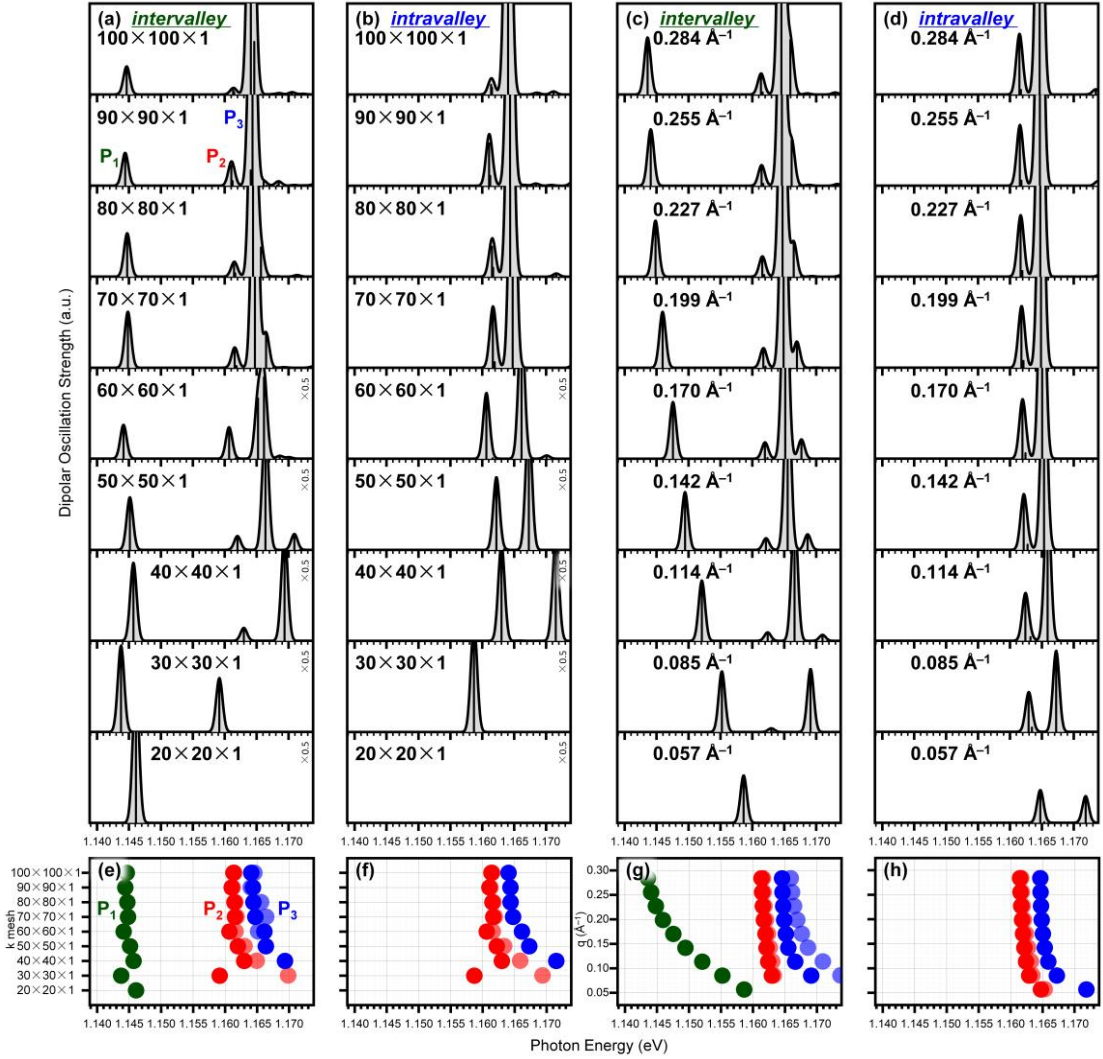

**Fig. 5 (a), (b), (c), (d),** Theoretical spectra calculated for ML-MoTe₂ based on the 3B-BSE for the intervalley **(a) & (c)** and intravalley **(b) & (d)** 3B cases, with respect to different calculation parameters, i.e. k-mesh density **(a) & (b)** and truncated radius around the K(K') point **(c) & (d)**. **(e), (f), (g), (h),** Plots of the spectral positions of P₁, P₂, & P₃ in **(a), (b), (c), & (d)**, respectively, as functions of the corresponding calculation parameters.

## Conclusion

The focus of this letter is to understand the true nature of the physical states and the corresponding spectral features of a three-body system of two electrons and one hole (or similarly one electron and two holes). This was done by both a theoretical calculation based on 3B-BSE combined with the cluster expansion technique and experimental measurement of photoluminescence spectroscopy under a gate-controlled charge background. Our consistent results from theories and experiments show the existence of three peaks, in contrast to the two-peak picture as commonly presented in theory or experiment for similar systems. Even though the full solution of the 3B-BSE was able to explain all the spectral features, the correspondence with the cluster expansion allowed us to identify the origin of the two peaks related to a 3B system, relating the two peaks to an irreducible cluster of △ and a product state of △ ~ △ , respectively.

Through the establishment of a systematic relationship between the Feynman diagrams of the 3B-BSE and cluster expansion technique, we showed that these new spectral features could be only explained by the inclusion of the 2e1h 3B irreducible cluster, which was necessary and sufficient in producing these features of $P_1 - P_3$. Through identifying the contributions to the spectrum of the intervalley and intravalley 3B states, we found the 3B irreducible entity, i.e. $P_1$, could only exist for the intervalley case but not for the intravalley one, while the two configurations of the weak coupling of an exciton and an electron, i.e. $P_2$ & $P_3$, could exist for both of above two cases.

Another important aspect of this paper relates to the terminology and physics of trion and charged exciton. Our consistent theory-experimental comparison showed that the charged exciton state of △ ~ △ is physically and spectrally different from the trion or triplon state of △ and their spectral splitting is of pure many-body origin. We believe that our results would contribute in a very important way to the fundamental understanding of the few-body physics. A similar few-body phenomenon was recently discovered as well for a 2e2h 4B system in ML-MoTe$_2$[18]. Also, such excited states and the related optical transitions would open the door to studying possible optical gain[7,41] and other non-linear properties, etc, in these systems as well.

**Data availability**
The data that support the findings of this study are available from the corresponding author upon request.

## References

1      Kraemer, T. *et al.* Evidence for Efimov quantum states in an ultracold gas of caesium atoms. *Nature* **440**, 315-318 (2006).

2      von Stecher, J., D'Incao, J. P. & Greene, C. H. Signatures of universal four-body phenomena and their relation to the Efimov effect. *Nature Physics* **5**, 417-421 (2009).

3      Mak, K. F. *et al.* Tightly bound trions in monolayer $MoS_2$. *Nature Materials* **12**, 207-211 (2013).

4      Ross, J. S. *et al.* Electrical control of neutral and charged excitons in a monolayer semiconductor. *Nat. Commun.* **4**, 1-6 (2013).

5      Yang, J. *et al.* Robust excitons and trions in monolayer $MoTe_2$. *ACS Nano* **9**, 6603-6609 (2015).

6      Plechinger, G. *et al.* Trion fine structure and coupled spin–valley dynamics in monolayer tungsten disulfide. *Nature Communications* **7**, 1-9 (2016).

7      Wang, Z. *et al.* Excitonic complexes and optical gain in two-dimensional molybdenum ditelluride well below the Mott transition. *Light: Science & Applications* **9**, 1-10 (2020).

8      Arora, A. *et al.* Dark trions govern the temperature-dependent optical absorption and emission of doped atomically thin semiconductors. *Physical Review B* **101**, 241413 (2020).

9      Zhang, Q. *et al.* Prolonging valley polarization lifetime through gate-controlled exciton-to-trion conversion in monolayer molybdenum ditelluride. *Nature Communications* **13**, 1-9 (2022).

10    Wang, Z. *et al.* Threshold-like Superlinear Accumulation of Excitons in a Gated Monolayer Transition Metal Dichalcogenide. *ACS Photonics* **10**, 412-420 (2023).

11    You, Y. *et al.* Observation of biexcitons in monolayer $WSe_2$. *Nature Physics* **11**, 477-481 (2015).

12    Shang, J. *et al.* Observation of excitonic fine structure in a 2D transition-metal dichalcogenide semiconductor. *ACS Nano* **9**, 647-655 (2015).

13    Plechinger, G. *et al.* Identification of excitons, trions and biexcitons in single-layer $WS_2$. *physica status solidi (RRL)–Rapid Research Letters* **9**, 457-461 (2015).

14    Sie, E. J., Frenzel, A. J., Lee, Y.-H., Kong, J. & Gedik, N. Intervalley biexcitons and many-body effects in monolayer $MoS_2$. *Physical Review B* **92**, 125417 (2015).

15    Steinhoff, A. *et al.* Biexciton fine structure in monolayer transition metal dichalcogenides. *Nature Physics* **14**, 1199-1204 (2018).

16    Nagler, P. *et al.* Zeeman splitting and inverted polarization of biexciton emission in monolayer $WS_2$. *Physical Review Letters* **121**, 057402 (2018).

17    Quang, N. H., Huong, N. Q., Dung, T. A., Tuan, H. A. & Thang, N. T. Strongly confined 2D parabolic quantum dot: Biexciton or quadron? *Physica B: Condensed Matter* **602**, 412591 (2021).

18    Tang, J. *et al.* The Quadruplon in a Monolayer Semiconductor. *arXiv preprint arXiv:2207.12760* (2022).

19    Rausch, R. & Potthoff, M. Multiplons in the two-hole excitation spectra of the one-dimensional Hubbard model. *New Journal of Physics* **18**, 023033 (2016).

20    Quang, N. H. & Huong, N. Q. Charged excitons and trions in 2D parabolic quantum dots. *Physica B: Condensed Matter* **633**, 413781 (2022).

21    Drüppel, M., Deilmann, T., Krüger, P. & Rohlfing, M. Diversity of trion states and substrate

## References

1      Kraemer, T. *et al.* Evidence for Efimov quantum states in an ultracold gas of caesium atoms. *Nature* **440**, 315-318 (2006).

2      von Stecher, J., D'Incao, J. P. & Greene, C. H. Signatures of universal four-body phenomena and their relation to the Efimov effect. *Nature Physics* **5**, 417-421 (2009).

3      Mak, K. F. *et al.* Tightly bound trions in monolayer $MoS_2$. *Nature Materials* **12**, 207-211 (2013).

4      Ross, J. S. *et al.* Electrical control of neutral and charged excitons in a monolayer semiconductor. *Nat. Commun.* **4**, 1-6 (2013).

5      Yang, J. *et al.* Robust excitons and trions in monolayer $MoTe_2$. *ACS Nano* **9**, 6603-6609 (2015).

6      Plechinger, G. *et al.* Trion fine structure and coupled spin–valley dynamics in monolayer tungsten disulfide. *Nature Communications* **7**, 1-9 (2016).

7      Wang, Z. *et al.* Excitonic complexes and optical gain in two-dimensional molybdenum ditelluride well below the Mott transition. *Light: Science & Applications* **9**, 1-10 (2020).

8      Arora, A. *et al.* Dark trions govern the temperature-dependent optical absorption and emission of doped atomically thin semiconductors. *Physical Review B* **101**, 241413 (2020).

9      Zhang, Q. *et al.* Prolonging valley polarization lifetime through gate-controlled exciton-to-trion conversion in monolayer molybdenum ditelluride. *Nature Communications* **13**, 1-9 (2022).

10    Wang, Z. *et al.* Threshold-like Superlinear Accumulation of Excitons in a Gated Monolayer Transition Metal Dichalcogenide. *ACS Photonics* **10**, 412-420 (2023).

11    You, Y. *et al.* Observation of biexcitons in monolayer $WSe_2$. *Nature Physics* **11**, 477-481 (2015).

12    Shang, J. *et al.* Observation of excitonic fine structure in a 2D transition-metal dichalcogenide semiconductor. *ACS Nano* **9**, 647-655 (2015).

13    Plechinger, G. *et al.* Identification of excitons, trions and biexcitons in single-layer $WS_2$. *physica status solidi (RRL)–Rapid Research Letters* **9**, 457-461 (2015).

14    Sie, E. J., Frenzel, A. J., Lee, Y.-H., Kong, J. & Gedik, N. Intervalley biexcitons and many-body effects in monolayer $MoS_2$. *Physical Review B* **92**, 125417 (2015).

15    Steinhoff, A. *et al.* Biexciton fine structure in monolayer transition metal dichalcogenides. *Nature Physics* **14**, 1199-1204 (2018).

16    Nagler, P. *et al.* Zeeman splitting and inverted polarization of biexciton emission in monolayer $WS_2$. *Physical Review Letters* **121**, 057402 (2018).

17    Quang, N. H., Huong, N. Q., Dung, T. A., Tuan, H. A. & Thang, N. T. Strongly confined 2D parabolic quantum dot: Biexciton or quadron? *Physica B: Condensed Matter* **602**, 412591 (2021).

18    Tang, J. *et al.* The Quadruplon in a Monolayer Semiconductor. *arXiv preprint arXiv:2207.12760* (2022).

19    Rausch, R. & Potthoff, M. Multiplons in the two-hole excitation spectra of the one-dimensional Hubbard model. *New Journal of Physics* **18**, 023033 (2016).

20    Quang, N. H. & Huong, N. Q. Charged excitons and trions in 2D parabolic quantum dots. *Physica B: Condensed Matter* **633**, 413781 (2022).

21    Drüppel, M., Deilmann, T., Krüger, P. & Rohlfing, M. Diversity of trion states and substrate

effects in the optical properties of an MoS$_2$ monolayer. *Nature Communications* **8**, 1-7 (2017).

22    Torche, A. & Bester, G. First-principles many-body theory for charged and neutral excitations: Trion fine structure splitting in transition metal dichalcogenides. *Physical Review B* **100**, 201403 (2019).

23    Klein, J. *et al.* Trions in MoS$_2$ are quantum superpositions of intra-and intervalley spin states. *Physical Review B* **105**, L041302 (2022).

24    Zhang, C., Wang, H., Chan, W., Manolatou, C. & Rana, F. Absorption of light by excitons and trions in monolayers of metal dichalcogenide MoS$_2$: Experiments and theory. *Physical Review B* **89**, 205436 (2014).

25    Florian, M. *et al.* The dielectric impact of layer distances on exciton and trion binding energies in van der Waals heterostructures. *Nano Letters* **18**, 2725-2732 (2018).

26    Tempelaar, R. & Berkelbach, T. C. Many-body simulation of two-dimensional electronic spectroscopy of excitons and trions in monolayer transition metal dichalcogenides. *Nature Communications* **10**, 1-7 (2019).

27    Zhumagulov, Y., Vagov, A., Senkevich, N. Y., Gulevich, D. & Perebeinos, V. Three-particle states and brightening of intervalley excitons in a doped MoS$_2$ monolayer. *Physical Review B* **101**, 245433 (2020).

28    Kudlis, A. & Iorsh, I. Modeling excitonic Mott transitions in two-dimensional semiconductors. *Physical Review B* **103**, 115307 (2021).

29    Salpeter, E. E. & Bethe, H. A. A relativistic equation for bound-state problems. *Physical Review* **84**, 1232 (1951).

30    Rohlfing, M. & Louie, S. G. Electron-hole excitations in semiconductors and insulators. *Physical Review Letters* **81**, 2312 (1998).

31    Rohlfing, M. & Louie, S. G. Electron-hole excitations and optical spectra from first principles. *Physical Review B* **62**, 4927 (2000).

32    Salij, A. & Tempelaar, R. Microscopic theory of cavity-confined monolayer semiconductors: Polariton-induced valley relaxation and the prospect of enhancing and controlling valley pseudospin by chiral strong coupling. *Physical Review B* **103**, 035431 (2021).

33    Mahan, G. D. *Many-particle physics*.    (Springer Science & Business Media, 2000).

34    Fetter, A. L. & Walecka, J. D. *Quantum theory of many-particle systems*.    (Courier Corporation, 2012).

35    Kira, M. & Koch, S. W. Many-body correlations and excitonic effects in semiconductor spectroscopy. *Progress in Quantum Electronics* **30**, 155-296 (2006).

36    Kira, M. & Koch, S. W. *Semiconductor quantum optics*.    (Cambridge University Press, 2011).

37    Almand-Hunter, A. *et al.* Quantum droplets of electrons and holes. *Nature* **506**, 471-475 (2014).

38    Liu, E. *et al.* Excitonic and valley-polarization signatures of fractional correlated electronic phases in a WSe$_2$/WS$_2$ moiré superlattice. *Physical Review Letters* **127**, 037402 (2021).

39    Sidler, M. *et al.* Fermi polaron-polaritons in charge-tunable atomically thin semiconductors. *Nature Physics* **13**, 255-261 (2017).

40    Efimkin, D. K. & MacDonald, A. H. Many-body theory of trion absorption features in two-dimensional semiconductors. *Physical Review B* **95**, 035417 (2017).

41      Hayamizu, Y. *et al.* Biexciton gain and the Mott transition in GaAs quantum wires. *Physical Review Letters* **99**, 167403 (2007).

**Acknowledgements**
J.T. thanks Qiyao Zhang and Jinhua Wu for their assistance with the cryostat. The authors acknowledge the following financial support: National Natural Science Foundation of China (Grant No. 91750206, No. 61861136006); Pingshan Innovation Platform Project of Shenzhen Hi-tech Zone Development Special Plan in 2022(29853M-KCJ-2023-002-01); Universities Engineering Technology Center of Guangdong (2023GCZX005); Key Programs Development Project of Guangdong (2022ZDJS111); Natural Science Foundation of Top Talent at SZTU (GDRC202301).

Corresponding authors
Correspondence to Cun-Zheng Ning, ningcunzheng@sztu.edu.cn

**Ethics declarations**
Competing interests
The authors declare no competing interests.

**Extended data figures and tables**